\documentclass[aps,prl,twocolumn,superscriptaddress,showpacs,preprintnumbers,amsmath,amssymb,tightenlines,columnwise,mlineno]{revtex4}

\usepackage{graphicx} 
\usepackage{siunitx}
\usepackage{xspace}
\usepackage[import,color,all]{xy}
\usepackage{relsize}

\newcommand{\BaBar}{\mbox{\textsl{B\hspace{-0.08em}{\smaller A}\hspace{-0.1em}B\hspace{-0.08em}{\smaller A\hspace{-0.12em}R}}}\xspace}
\def\bpiplnu{\ensuremath{\bar{B}^{0}\!\rightarrow\!\pi^+\ell^{-}\bar{\nu}_\ell}\xspace}
\def\bomegalnu{\ensuremath{B^{-}\rightarrow\omega\ell^{-}\bar{\nu}_\ell}\xspace}
\def\bxulnu{\ensuremath{\bar{B}\rightarrow X_u\ell^{-}\bar{\nu}_\ell}\xspace}
\def\xpulnu{\ensuremath{X_u^{+}\ell^{-}\bar{\nu}_\ell}\xspace}
\def\xzulnu{\ensuremath{X_u^{0}\ell^{-}\bar{\nu}_\ell}\xspace}
\def\bbpilmnub{\ensuremath{\bar{B}\rightarrow \pi\ell^{-}\bar{\nu}_\ell}\xspace}
\def\brholnu{\ensuremath{\bar{B}\rightarrow \rho\ell^{-}\bar{\nu}_\ell}\xspace}
\def\bb{\ensuremath{{B\bar{B}}}\xspace}
\def\bpbm{\ensuremath{{B^+B^-}}\xspace}
\def\qsq{\ensuremath{q^2}\xspace}
\def\GeVc{GeV\!/\!\ensuremath{{\it c}}\xspace}
\def\GeVcc{GeV\!/\!\ensuremath{{\it c}^2}\xspace}
\def\MeVc{MeV\!/\!\ensuremath{{\it c}}\xspace}
\def\mbc{\ensuremath{M_\mathrm{bc}}\xspace}
\def\Br{{\cal B}}
\def\nnout{\ensuremath{o_\mathrm{nn}}\xspace}
\def\vub{\ensuremath{|V_{ub}|}\xspace}
\def\bmmumnub{\ensuremath{B^{-}\to\mu^{-}\bar\nu_\mu}\xspace}
\def\bmunu{\ensuremath{B\to\mu\bar\nu_\mu}\xspace}
\def\bmtaumnub{\ensuremath{B^{-} \to \tau^{-} \bar\nu_\tau}\xspace}
\def\bmemnub{\ensuremath{B^{-} \to e^{-} \bar\nu_e}\xspace}
\def\bmellmnub{\ensuremath{B^{-}\to\ell^{-}\bar\nu_\ell}\xspace}
\def\bpilnu{\ensuremath{B\to\pi\mu\bar{\nu}_\mu}\xspace}
\def\pvi{\ensuremath{\vec{\mathbf{p}}^{*}_{i}}\xspace}

\def\IntLumShort{\ensuremath{711\ \mathrm{fb}^{-1}}\!\xspace}
\def\Nbb{\ensuremath{772 \times 10^6} \bb pairs\xspace}
\def\NbbSVDOne{\ensuremath{152 \times 10^6} \bb pairs\xspace}
\def\NbbSVDTwo{\ensuremath{620 \times 10^6} \bb pairs\xspace}
\def\Nbbwithshorterr{\ensuremath{\left( 772 \pm 11 \right) \times 10^6}\xspace}

\begin{document}
\vspace*{-3\baselineskip}
\resizebox{!}{3cm}{\includegraphics{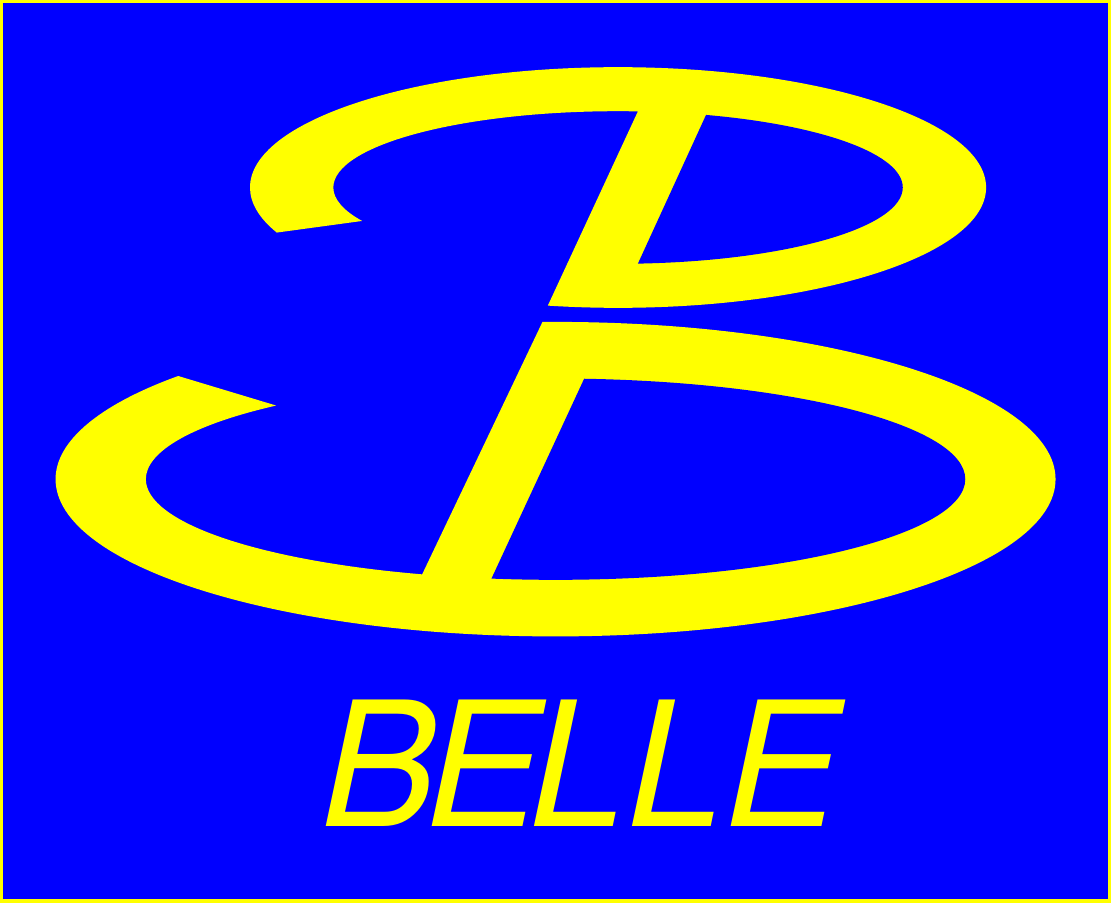}}

\preprint{
\vbox{\hbox{}
   \hbox{Belle Preprint 2017-17}
   \hbox{KEK Preprint 2017-21}
}}
  
\title{\boldmath Search for \bmmumnub Decays at the Belle Experiment}

\noaffiliation
\affiliation{University of the Basque Country UPV/EHU, 48080 Bilbao}
\affiliation{Beihang University, Beijing 100191}
\affiliation{Budker Institute of Nuclear Physics SB RAS, Novosibirsk 630090}
\affiliation{Faculty of Mathematics and Physics, Charles University, 121 16 Prague}
\affiliation{Chonnam National University, Kwangju 660-701}
\affiliation{University of Cincinnati, Cincinnati, Ohio 45221}
\affiliation{Deutsches Elektronen--Synchrotron, 22607 Hamburg}
\affiliation{University of Florida, Gainesville, Florida 32611}
\affiliation{Justus-Liebig-Universit\"at Gie\ss{}en, 35392 Gie\ss{}en}
\affiliation{Gifu University, Gifu 501-1193}
\affiliation{SOKENDAI (The Graduate University for Advanced Studies), Hayama 240-0193}
\affiliation{Gyeongsang National University, Chinju 660-701}
\affiliation{Hanyang University, Seoul 133-791}
\affiliation{University of Hawaii, Honolulu, Hawaii 96822}
\affiliation{High Energy Accelerator Research Organization (KEK), Tsukuba 305-0801}
\affiliation{J-PARC Branch, KEK Theory Center, High Energy Accelerator Research Organization (KEK), Tsukuba 305-0801}
\affiliation{IKERBASQUE, Basque Foundation for Science, 48013 Bilbao}
\affiliation{Indian Institute of Technology Bhubaneswar, Satya Nagar 751007}
\affiliation{Indian Institute of Technology Guwahati, Assam 781039}
\affiliation{Indian Institute of Technology Madras, Chennai 600036}
\affiliation{Indiana University, Bloomington, Indiana 47408}
\affiliation{Institute of High Energy Physics, Chinese Academy of Sciences, Beijing 100049}
\affiliation{Institute of High Energy Physics, Vienna 1050}
\affiliation{Institute for High Energy Physics, Protvino 142281}
\affiliation{INFN - Sezione di Napoli, 80126 Napoli}
\affiliation{INFN - Sezione di Torino, 10125 Torino}
\affiliation{Advanced Science Research Center, Japan Atomic Energy Agency, Naka 319-1195}
\affiliation{J. Stefan Institute, 1000 Ljubljana}
\affiliation{Kanagawa University, Yokohama 221-8686}
\affiliation{Institut f\"ur Experimentelle Kernphysik, Karlsruher Institut f\"ur Technologie, 76131 Karlsruhe}
\affiliation{Kennesaw State University, Kennesaw, Georgia 30144}
\affiliation{King Abdulaziz City for Science and Technology, Riyadh 11442}
\affiliation{Department of Physics, Faculty of Science, King Abdulaziz University, Jeddah 21589}
\affiliation{Korea Institute of Science and Technology Information, Daejeon 305-806}
\affiliation{Korea University, Seoul 136-713}
\affiliation{Kyoto University, Kyoto 606-8502}
\affiliation{Kyungpook National University, Daegu 702-701}
\affiliation{\'Ecole Polytechnique F\'ed\'erale de Lausanne (EPFL), Lausanne 1015}
\affiliation{P.N. Lebedev Physical Institute of the Russian Academy of Sciences, Moscow 119991}
\affiliation{Faculty of Mathematics and Physics, University of Ljubljana, 1000 Ljubljana}
\affiliation{Ludwig Maximilians University, 80539 Munich}
\affiliation{Luther College, Decorah, Iowa 52101}
\affiliation{University of Malaya, 50603 Kuala Lumpur}
\affiliation{University of Maribor, 2000 Maribor}
\affiliation{Max-Planck-Institut f\"ur Physik, 80805 M\"unchen}
\affiliation{School of Physics, University of Melbourne, Victoria 3010}
\affiliation{University of Miyazaki, Miyazaki 889-2192}
\affiliation{Moscow Physical Engineering Institute, Moscow 115409}
\affiliation{Moscow Institute of Physics and Technology, Moscow Region 141700}
\affiliation{Graduate School of Science, Nagoya University, Nagoya 464-8602}
\affiliation{Kobayashi-Maskawa Institute, Nagoya University, Nagoya 464-8602}
\affiliation{Nara Women's University, Nara 630-8506}
\affiliation{National Central University, Chung-li 32054}
\affiliation{National United University, Miao Li 36003}
\affiliation{Department of Physics, National Taiwan University, Taipei 10617}
\affiliation{H. Niewodniczanski Institute of Nuclear Physics, Krakow 31-342}
\affiliation{Nippon Dental University, Niigata 951-8580}
\affiliation{Niigata University, Niigata 950-2181}
\affiliation{Novosibirsk State University, Novosibirsk 630090}
\affiliation{Osaka City University, Osaka 558-8585}
\affiliation{Pacific Northwest National Laboratory, Richland, Washington 99352}
\affiliation{University of Pittsburgh, Pittsburgh, Pennsylvania 15260}
\affiliation{Punjab Agricultural University, Ludhiana 141004}
\affiliation{Theoretical Research Division, Nishina Center, RIKEN, Saitama 351-0198}
\affiliation{University of Science and Technology of China, Hefei 230026}
\affiliation{Showa Pharmaceutical University, Tokyo 194-8543}
\affiliation{Soongsil University, Seoul 156-743}
\affiliation{Stefan Meyer Institute for Subatomic Physics, Vienna 1090}
\affiliation{Sungkyunkwan University, Suwon 440-746}
\affiliation{School of Physics, University of Sydney, New South Wales 2006}
\affiliation{Department of Physics, Faculty of Science, University of Tabuk, Tabuk 71451}
\affiliation{Tata Institute of Fundamental Research, Mumbai 400005}
\affiliation{Excellence Cluster Universe, Technische Universit\"at M\"unchen, 85748 Garching}
\affiliation{Department of Physics, Technische Universit\"at M\"unchen, 85748 Garching}
\affiliation{Toho University, Funabashi 274-8510}
\affiliation{Department of Physics, Tohoku University, Sendai 980-8578}
\affiliation{Earthquake Research Institute, University of Tokyo, Tokyo 113-0032}
\affiliation{Department of Physics, University of Tokyo, Tokyo 113-0033}
\affiliation{Tokyo Institute of Technology, Tokyo 152-8550}
\affiliation{Tokyo Metropolitan University, Tokyo 192-0397}
\affiliation{University of Torino, 10124 Torino}
\affiliation{Virginia Polytechnic Institute and State University, Blacksburg, Virginia 24061}
\affiliation{Wayne State University, Detroit, Michigan 48202}
\affiliation{Yamagata University, Yamagata 990-8560}
\affiliation{Yonsei University, Seoul 120-749}
  \author{A.~Sibidanov}\altaffiliation[now at ]{University of Victoria, Victoria, British Columbia, Canada V8W 3P6}{\affiliation{School of Physics, University of Sydney, New South Wales 2006}} 
  \author{K.~E.~Varvell}\affiliation{School of Physics, University of Sydney, New South Wales 2006} 
  \author{I.~Adachi}\affiliation{High Energy Accelerator Research Organization (KEK), Tsukuba 305-0801}\affiliation{SOKENDAI (The Graduate University for Advanced Studies), Hayama 240-0193} 
  \author{H.~Aihara}\affiliation{Department of Physics, University of Tokyo, Tokyo 113-0033} 
  \author{S.~Al~Said}\affiliation{Department of Physics, Faculty of Science, University of Tabuk, Tabuk 71451}\affiliation{Department of Physics, Faculty of Science, King Abdulaziz University, Jeddah 21589} 
  \author{D.~M.~Asner}\affiliation{Pacific Northwest National Laboratory, Richland, Washington 99352} 
  \author{T.~Aushev}\affiliation{Moscow Institute of Physics and Technology, Moscow Region 141700} 
  \author{R.~Ayad}\affiliation{Department of Physics, Faculty of Science, University of Tabuk, Tabuk 71451} 
  \author{V.~Babu}\affiliation{Tata Institute of Fundamental Research, Mumbai 400005} 
  \author{I.~Badhrees}\affiliation{Department of Physics, Faculty of Science, University of Tabuk, Tabuk 71451}\affiliation{King Abdulaziz City for Science and Technology, Riyadh 11442} 
% ...
\collaboration{The Belle Collaboration}

\begin{abstract}
  We report the result of a search for the decay \bmmumnub.  The
  signal events are selected based on the presence of a high momentum
  muon and the topology of the rest of the event showing properties of
  a generic $B$-meson decay, as well as the missing energy and
  momentum being consistent with the hypothesis of a neutrino from the
  signal decay.
  We find a 2.4 standard deviation excess above background including systematic
  uncertainties, which corresponds to a branching fraction of $\Br(\bmmumnub)
  =(6.46 \pm 2.22 \pm 1.60 )\times10^{-7}$ or a frequentist 90\% confidence
  level interval on the \bmmumnub branching fraction of $[2.9, 10.7]\times 10^{-7}$.
  This result is obtained
  from a \IntLumShort data sample that contains \Nbb, collected near the 
  $\Upsilon(4S)$ resonance with the Belle detector at the KEKB asymmetric-energy
  $e^+ e^-$ collider.
\end{abstract}

\pacs{13.20.-v, 14.40.Nd, 12.15.Hh, 12.38.Gc}

\maketitle

In the Standard Model (SM), the branching fraction for the
purely leptonic decay of a $B^-$ meson~\cite{conj}, assuming a massless neutrino, is:
\begin{equation}
\label{eq:SMBF}
\Br(\bmellmnub) =
\dfrac{G_{F}^{2}m_{B}m_{\ell}^2}{8\pi}\left(1-\dfrac{m_\ell^2}{m_B^2}\right)^2
f_B^2 \vub^2\tau_B, 
\end{equation} 
where $G_{F}$ is the Fermi constant, $m_B$ and $m_\ell$ are the masses
of the $B$ meson and charged lepton, respectively, $f_B$ is the
$B$-meson decay constant obtained from theory, $\tau_B$ is the
lifetime of the $B$ meson and $V_{ub}$ is the CKM matrix element
governing the coupling between $u$ and $b$ quarks.  The FLAG~\cite{Aoki:2016frl} average of
lattice QCD calculations gives $f_B =
0.186\pm0.004$ GeV, and the world-average value of $\tau_{B}$ is
$1.638\pm0.004$~ps~\cite{PDG}.  For the value of \vub, we repeat
the fit procedure described in Ref.~\cite{b2ulnu}, equipped with the most
recent lattice QCD calculation by the FNAL/MILC
collaborations~\cite{newlqcd2} that provides a tight constraint on the
hadronic form-factor $f_+(\qsq)$ governing exclusive \bpiplnu decays.
The form-factor parameters for \bpiplnu decay are also obtained with
this procedure.  The value of \vub thus obtained is
$\vub\times 10^3 = 3.736\pm0.142$ with fit quality
$\chi^2 = 47.9$ for 45 degrees of freedom.  Using these values as input parameters
for Eq.~\ref{eq:SMBF}, the expected branching fractions for
$\bmellmnub$ decays are displayed in
Table~\ref{table:lnuprediction}. Also shown in the Table are the
expected event yields for \bmellmnub decays in the full Belle
data set, where we use $\Br(\Upsilon(4S)\to\bpbm) =
0.514\pm0.006$~\cite{PDG}.

\begin{table}[tbh]
  \caption{\label{table:lnuprediction} The expected branching
    fractions and event yields in the full Belle data sample of
    $772\times10^6$ \bb events for the decay \bmellmnub.}  \centering
  \begin{tabular}{ccc}
    \hline
    ~~~$\ell$~~~ & $\Br_\mathrm{SM}$~~~ & $N_\mathrm{SM}^\mathrm{Belle}$ \\
    \hline
    $\tau$ & $( 8.45 \pm  0.70) \times 10^{ -5}$~~~  & $ (670 \pm 57) \times 10^2$ \\
    $ \mu$ & $( 3.80 \pm  0.31) \times 10^{ -7}$~~~  & $   301 \pm     25$ \\
    $   e$ & $( 8.89 \pm  0.73) \times 10^{-12}$~~~  & $ 0.0071 \pm 0.0006$ \\
    \hline
  \end{tabular}
\end{table}

Due to the relatively small theoretical uncertainties within the SM
framework, \bmellmnub decays are good candidates for testing SM
predictions and searching for phenomena that might modify them.  For
instance, the effects of charged Higgs bosons in two-Higgs-doublet models
of type-II~\cite{2HDM}, the R-parity-violating Minimal Supersymmetric Standard Model
(MSSM)~\cite{MSSM}, or leptoquarks~\cite{LQ} may significantly
change the \bmellmnub decay rates.

Moreover, by taking the ratios of purely leptonic $B^-$ decays, most
of the input parameters in Eq.~\ref{eq:SMBF} cancel and very
precise values are predicted. Predictions of the
ratios $\Br(\bmtaumnub)/\Br(\bmemnub)$ and
$\Br(\bmtaumnub)/\Br(\bmmumnub)$ obtained within a general MSSM
at large $\tan\beta$~\cite{tanbeta} with heavy squarks~\cite{Lepuniv}
deviate from the SM expectations and the deviation can be as large as an
order of magnitude in the grand unified theory framework~\cite{FI}.

There have been several searches for the decay \bmmumnub to
date~\cite{Yook:2014kga,Satoyama:2006xn,Aubert:2009ar,Aubert:2008cu,Aubert:2009wt}
and no evidence of the decay has been found, with the most stringent
limit of $\Br(\bmmumnub)<1.0\times10^{-6}$ at 90\% confidence level
set by the \BaBar collaboration using an untagged
method~\cite{Aubert:2009ar}.

In this article, we present a search for the decay \bmmumnub that also uses 
the untagged method. This study is based on a \IntLumShort data
sample that contains \Nbbwithshorterr \bb pairs, collected with the
Belle detector at the KEKB asymmetric-energy $e^+e^-$ (3.5 on 8~GeV)
collider~\cite{KEKB} operating at the $\Upsilon(4S)$ resonance.

The Belle detector is a large-solid-angle magnetic spectrometer that
consists of a silicon vertex detector (SVD), a 50-layer central drift
chamber (CDC), an array of aerogel threshold Cherenkov counters (ACC),
a barrel-like arrangement of time-of-flight scintillation counters
(TOF) and an electromagnetic calorimeter comprised of CsI(Tl)
crystals (ECL) located inside a superconducting solenoid coil that
provides a 1.5~T magnetic field.  An iron flux-return yoke located outside
of the coil is instrumented to detect $K_L^0$ mesons and to identify
muons (KLM).  The detector is described in detail
elsewhere~\cite{Belle}.  Two inner detector configurations were
used. A 2.0 cm beampipe and a 3-layer silicon vertex detector were
used for the first sample of \NbbSVDOne, while a 1.5 cm beampipe, a
4-layer silicon detector and a small-cell inner drift chamber were
used to record the remaining \NbbSVDTwo~\cite{svd2}.

The data were collected at a center-of-mass energy of \SI{10.58}{\giga\electronvolt}, corresponding to
the $\Upsilon(4S)$ resonance.
The size of the data sample is equivalent to an integrated luminosity
of \IntLumShort. We also utilise a sample of 79 fb$^{-1}$ collected
below the \bb threshold to characterize the contribution of the $e^+e^-\to
q\bar{q}$ process, so-called continuum, where $q$ is either a $u$,
$d$, $s$, or $c$ quark; this is one of the major
backgrounds.

We use Monte Carlo~(MC) samples based on the detailed detector
geometry description implemented with the {\tt GEANT3} package
\cite{GEANT3} to establish the analysis technique and study major
backgrounds. Events with $B$-meson decays are generated using {\tt
  EvtGen}~\cite{evtgen}. The generated samples include $2\times10^6$
signal events, a sample of generic \bb decays corresponding to ten
times the integrated luminosity of the data, continuum corresponding
to six times the data, \bxulnu decays corresponding to twenty times the
data, other $B$ decays with probability $\lesssim 4\times10^{-4}$
corresponding to fifty times the data, and $e^+e^-\to \tau^+\tau^-$
corresponding to five times the data, as well as other QED and
two-photon processes with various multiples of the data. 
The simulation accounts for the evolution in background conditions and beam
collision parameters. Final-state radiation from charged particles is
modelled using the PHOTOS package~\cite{photos}.

MC samples for one of the largest backgrounds from $B$ decays,
charmless semileptonic decays, are generated according to the number of
\bb pairs in data, scaled 20 times, assuming inclusive semileptonic branching 
fractions of $\Br(\bar{B}^0\to\xpulnu) = 1.709 \times 10^{-3}$ and
$\Br(B^-\to\xzulnu) = 1.835 \times 10^{-3}$.
Samples with \bbpilmnub, \brholnu, and \bomegalnu decays are modelled using
Light Cone Sum Rule form-factor
predictions~\cite{ball01,ball05}.  Other decays to exclusive meson
states are modelled using the updated quark model by
Isgur-Scora-Grinstein-Wise~\cite{ISGW2}. The inclusive
component of charmless semileptonic decays is modelled to leading order
in $\alpha_s$ based on a prediction in the Heavy-Quark Expansion
framework~\cite{DeFazio}. The fragmentation process of the resulting
parton to the final hadron state is modelled using the PYTHIA6.2
package~\cite{pythia62}.

In addition, $8\times 10^6$ \bbpilmnub MC events are generated
uniformly as a function of \qsq. These events are reweighted to the
most recent lattice QCD form-factor calculation, in order to decrease
MC statistical fluctuations at high \qsq and to study the behavior 
of the fit procedure described below when form-factors are varied 
within uncertainties.

Finally, $10^6$ events of the three-body decay $B^{-}\to\mu^{-}\bar\nu_\mu\gamma$
are generated with photon energy above \SI{25}{\mega\electronvolt} in the $B$ decay frame
with the form-factor parameters $R=3$ and $m_b=\SI{5}{\giga\electronvolt}$ based on the
work in Ref.~\cite{lnuggen}.

The muon in \bmmumnub decay is monochromatic in the absence of
radiation, with an energy of half the $B$-meson rest mass energy in the
$B$-meson rest frame. In the $\Upsilon(4S)$ center-of-mass frame, where the $B$ meson
is in motion, the boost smears the momentum of
the muon, $p_\mu^*$, to the range (2.476, 2.812)~$\SI{}{\GeVc}$. We
select well-reconstructed muon candidates in the wider region of (2.2,
4.0)~$\SI{}{\GeVc}$ to include enough data to validate
the analysis procedure and estimate backgrounds. A blind analysis is
performed with the $\Upsilon(4S)$ data in the $p_\mu^*$ interval
(2.45, 2.85)~$\SI{}{\GeVc}$ excluded until the analysis procedure has
been finalized.
Signal muons are identified by a standard procedure based on their
penetration range and degree of transverse scattering in the KLM
detector with an efficiency of $\sim90$\%~\cite{muid}. An additional
selection is applied with information from the CDC, ECL, ACC, and TOF
subdetectors, combined using an artificial neural network, to reject the charged-kaon muonic
decay in flight. Background suppression of 33\% is achieved by
this procedure, with a signal-muon selection efficiency of 97\%.

Charged particles, including the signal muon candidate, are required to originate from the region near
the interaction point (IP) of the electron and positron beams.
This region is defined by $|z_{\rm PCA}| < 2\ \rm cm$ and $r_{\rm PCA}
< 0.5\ \rm cm$, where $z_{\rm PCA}$ is the distance of the point of closest approach (PCA) from the IP along
the $z$ axis (opposite the positron beam) and $r_{\rm PCA}$ is the
distance from this axis in the transverse plane.
The charged daughters of reconstructed long-lived neutral particles
(converted $\gamma$, $K_S^0$, and $\Lambda$) are included in this list
even if they fail the IP selection.
All other charged particles
are ignored. We discard the event if the total momentum of these
particles exceeds \SI{1.3}{\GeVc} to suppress
the background from mis-reconstructed long-lived neutral particles.

Each surviving track that is not classified as a long-lived
neutral-particle daughter is assigned a unique identity.  Electrons
are identified using the ratio of the energy detected in the ECL to
the track momentum, the ECL shower shape, position matching between
the track and ECL cluster, the energy loss in the CDC, and the
response of the ACC~\cite{eid}. Muons are identified as described
earlier for the signal muon candidates. Pions, kaons and protons are
identified using the responses of the CDC, ACC, and TOF. In the
expected momentum region for particles from $B$-meson decays, charged
leptons are identified with an efficiency of about 75\% while the
probability to misidentify a pion as an electron (muon) is 1.9\%
(5\%). Charged pions (kaons, protons) are selected with an efficiency
of 86\% (75\%, 98\%) and a pion (kaon, proton) misidentification
probability of 6\% (13\%, 72\%).

Photon candidates are selected using a polar-angle-dependent energy
threshold chosen such that a photon with energy above (below) the
threshold is more likely to originate from $B$-meson decay
(calorimeter noise).  In the barrel calorimeter, the energy threshold
is about \SI{40}{\mega\electronvolt}; in the forward and backward
endcaps, it rises to \SI{110}{\mega\electronvolt} and
\SI{150}{\mega\electronvolt}, respectively.  Additionally, we require
the total energy deposition in the calorimeter not associated with
charged particles nor recognized as photons to be under
$\SI{0.6}{\giga\electronvolt}$.

The neutrino in \bmmumnub decay is not detected. The photons and surviving charged particles 
other than the signal muon should come from the
companion $B$ meson in the $e^+e^-\to\Upsilon(4S)\to B^+B^-$ process. We select companion $B$ meson candidates that
have invariant mass close to the nominal $B$-meson mass and total energy
close to the nominal $B$-meson energy from the $\Upsilon(4S)\to\bb$ decay. These
quantities are represented by the beam-constrained mass and energy 
\begin{eqnarray}
\mbc & = & \sqrt{E_\mathrm{beam}^2/c^4 - |\sum_i \pvi/c|^2},\\
E_B & = & \sum_i\sqrt{(m_ic^2)^2+|\pvi c|^2},
\end{eqnarray}
where $E_\mathrm{beam}$ is the beam energy in the $\Upsilon(4S)$
center-of-mass frame, and \pvi and $m_i$ are
the center-of-mass frame momentum and mass, respectively, of the $i^{\mathrm{th}}$
particle that makes up the accompanying $B$-meson candidate.
We retain events that satisfy 
$\mbc>\SI{5.1}{\GeVcc}$ and
$\SI{-3}{\GeV}<E_B-E_\mathrm{beam}<\SI{2}{GeV}$.

To exploit the jet-like structure of non-\bb background, where
particles tend to be produced collinearly, we define the direction $\hat{n}$ of the thrust axis by maximizing the quantity
\begin{equation}
  \frac{\sum\limits_i (\hat{n}\cdot \pvi)^2}{\sum\limits_i |\pvi|^2},
  \label{thrust}
\end{equation}
while satisfying the condition $\hat{n} \cdot ( \sum\limits_i \pvi ) >
0$. We require $\hat{n}\cdot\hat{p}_{\mu}^{*} > -0.8$, where
$\hat{p}{}_{\mu}^{*}$ is the signal-muon direction, to remove muons
collinear with the other particles in the event.

The missing energy of a neutrino from semileptonic decays of $B$ or
$D$ mesons can be similar to that of the signal, and an excess of
reconstructed charged leptons is a signature of these decays. We
therefore require no more than one additional lepton in the
event besides the signal muon.

The information from the KLM detector subsystem is also used to
improve signal purity.  We require no more than one $K_L^0$ cluster in
the KLM and no $K_L^0$ clusters associated with ECL clusters. This
selection rejects about 24\% of background events and keeps about 90\%
of signal. The $K_L^0$ detection efficiency is calibrated using a
$D^0\to \phi K_S^0$ control sample.

The total signal selection efficiency for  \bmmumnub decays is estimated at 
this stage to be around 38\%, with an expected signal yield of $115 \pm 9$.

After all of the selections described above are applied, the remaining 
background is still more than three orders of magnitude larger than the 
expected signal yield.  A multivariate data analysis is employed to further 
separate signal from background.  We combine various kinematic parameters 
of an event into a single variable \nnout using an artificial 
neural network. We choose 14 input parameters 
that are uncorrelated with the absolute value of the muon momentum, and 
that collectively yield the best signal to background ratio. These parameters are
five event-shape moments, the polar angle of the missing momentum vector, the
angle between the thrust axis and the signal-muon direction, the energy 
difference $E_B-E_\text{beam}$, the angle between the signal-muon direction and
the thrust axis calculated using only photons, 
the angle between the momentum of the companion $B$ meson and the signal-muon direction, 
the $z$-axis distance between the signal muon's $z_\text{PCA}$ and the reconstructed
vertex of the companion $B$ meson, the square of the thrust as defined
in Eq.\ (\ref{thrust}), the sum of charges of charged particles in an event, 
and the polar angle of the muon momentum vector.

\begin{figure}
  \begin{xy}
    \xyimport(1,1)(0,0){\includegraphics[width=\linewidth]{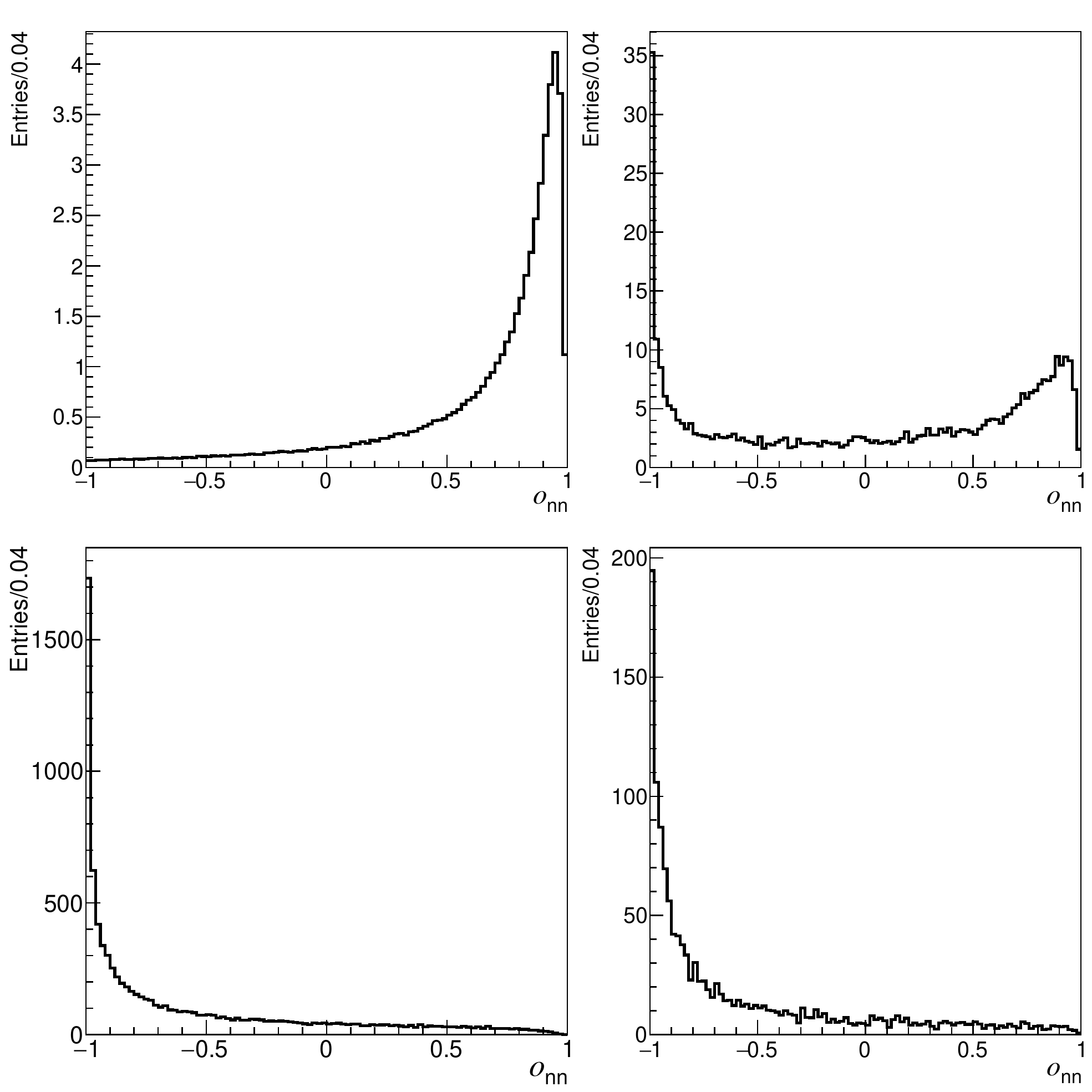}},
    (0.15,0.94)*!L\txt{\bmmumnub},
    (0.65,0.94)*!L\txt{\bbpilmnub},
    (0.15,0.47)*!L\txt{Continuum},
    (0.65,0.47)*!L\txt{$e^+e^-\to\tau^+\tau^-$}
  \end{xy}
  \caption{\label{fig:nnout}The distributions of the neural network output
    variable for the signal and major background processes predicted
    by MC in the signal-enhanced region $\SI{2.644}{\GeVc} < p_\mu^* <
    \SI{2.812}{\GeVc}$.}
\end{figure}

The employed configuration of the network consists of the input layer
and two hidden layers having 56 and 28 neurons and the $\tanh$
activation function; in total, it has 2465 parameters to optimize. The MC
sample is divided into equal training and testing parts with almost 2
million events in each. The distributions of the neural network output
variables in the signal-enhanced momentum region are shown in
Fig.~\ref{fig:nnout}. The only background components peaking in the
signal region are \bbpilmnub and, much less prominently, \brholnu. All
other major backgrounds decrease significantly approaching the $\nnout\sim 1$
region and do not have a peaking behavior in the $\nnout$ variable that can mimic the signal.

The signal yield is extracted by a binned maximum-likelihood fit in
the $p_\mu^*$-\nnout plane using the method described in 
Ref.~\cite{Barlow:1993dm}, taking into account the uncertainty arising from the
finite number of events in the template MC histograms. The fit region
covers muon momenta from \SI{2.2}{} to \SI{4}{\GeVc} with
\SI{50}{\MeVc} bins and the full range of the \nnout variable
from $-1$ to 1 with 0.04 bins. The region at high muon momentum
$p_\mu^*$ and high \nnout is sparsely populated; to avoid bins
with zero or a few events, which are undesirable for the fit method
employed, we increased the bin size in this region. The fine binning in
the signal region is preserved. After the rebinning, the
$p_\mu^*$-\nnout histogram is reduced from
1800 to 1226 bins. The fit method tends to scale
low-populated templates to improve the fit to data; because of this, background
components with the predicted fraction of under 1\% of the total number of events are fixed in the fit to
the MC prediction. The fitted-yield components are the signal, \bbpilmnub,
\brholnu, the rest of the charmless semileptonic decays, \bb,
$c\bar{c}$, $uds$, $\tau^+\tau^-$, and $e^+e^-\mu^+\mu^-$. The fixed-yield
components are $\mu^+\mu^-$, $e^+e^-e^+e^-$, $e^+e^-u\bar{u}$,
$e^+e^-s\bar{s}$, and $e^+e^-c\bar{c}$.

\begin{figure}
  \includegraphics[width=\linewidth]{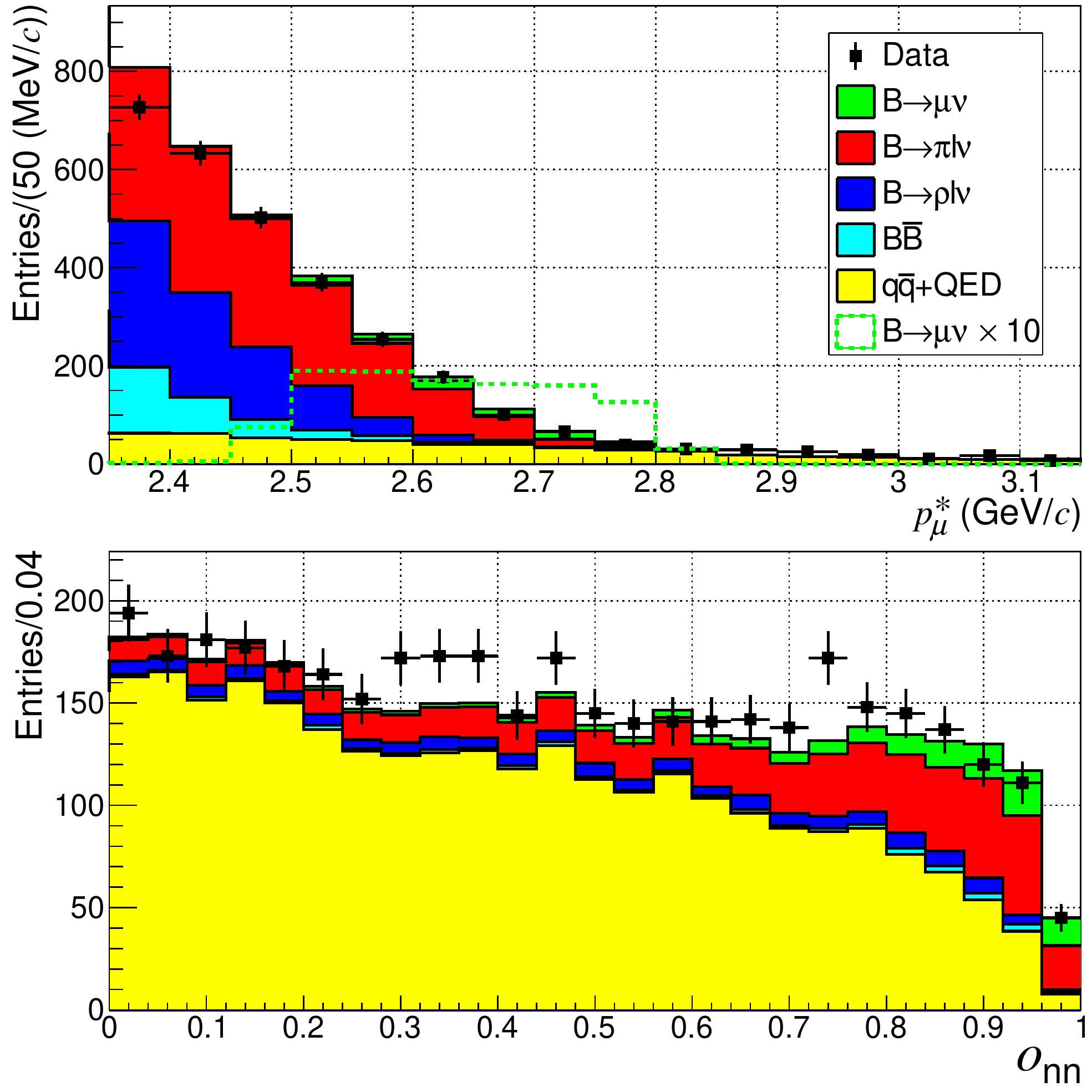}
  \caption{\label{fig:projs} Projections of the fitted distribution to
    data onto the histogram axes in the signal-enhanced regions
    $0.84<\nnout$ (top plot) and
    $\SI{2.6}{\GeVc}<p_\mu^*<\SI{2.85}{\GeVc}$ (bottom plot).  }
\end{figure}

To obtain the signal branching fraction, we fit the ratio $R =
N_{\bmunu}/N_{\bpilnu}$. This ratio also helps to reliably estimate the
fit uncertainty. The result of the fit is $R=(1.66 \pm
0.57 )\times10^{-2}$, which is equivalent to a signal yield of
$N_{\bmunu}=195\pm67$ and the branching fraction ratio of $\Br(\bmmumnub)/\Br(\bbpilmnub) =
(4.45 \pm 1.53_\text{stat}) \times 10^{-3}.$ This result can be compared to the MC
prediction of this ratio $R_\text{MC} =
114.6/11746=0.976\times10^{-2}$, obtained assuming
$\Br(\bmunu)=3.80\times10^{-7}$ and $\Br(\bbpilmnub) = 1.45\times10^{-4}$
(the PDG average~\cite{PDG}). The fitted value of $R$ results in the branching fraction
$\Br(\bmunu)=(6.46 \pm 2.22)\times10^{-7}$, where the quoted
uncertainty is statistical only. The statistical significance of the signal is
3.4$\sigma$, determined from the likelihood ratio of the fits with a free signal
component and with the signal component fixed to zero. The fit result of
the reference process \bbpilmnub agrees with the MC prediction to better than
10\%. The projections of the fitted distribution in the signal-enhanced
regions are shown in Fig.~\ref{fig:projs}.
The fit qualities of the displayed projections are
$\chi^2/\text{ndf} = 27.6/16$ (top panel) and $\chi^2/\text{ndf}
= 29.1/25$ (bottom panel), taking into account only data uncertainties.

The double ratio $R/R_\text{MC}$ benefits from substantial
cancellation of the systematic uncertainties from muon identification,
lepton and neutral-kaon vetos and the companion $B$-meson decay
mis-modelling, as well as partially cancelling trigger uncertainties
and possible differences in the distribution of the \nnout variable.

In the signal region, the main background contribution comes from
charmless semileptonic decays; in particular, the main
components \bbpilmnub and \brholnu, which peak at high \nnout values,
are carefully studied. With soft and undetected hadronic recoil, these
decays are kinematically indistinguishable from the signal in an
untagged analysis. For the
\bbpilmnub component, we vary the form-factor shape within
uncertainties obtained with the new lattice QCD result~\cite{newlqcd2} and
the procedure described in Ref.~\cite{b2ulnu}, which was used to estimate
the value of \vub. Since the form-factor is tightly constrained, the 
contribution to the systematic
uncertainty from the \bbpilmnub background is estimated to be only
0.9\%. For the \brholnu component, the form-factors at high
\qsq or high muon momentum have much larger uncertainties and several
available calculations are employed~\cite{ball05,ISGW2,ukqcd},
resulting in a systematic uncertainty of 12\%.

The rare hadronic decay $B^-\to K_L^0 \pi^-$, where $K_L^0$ is not
detected and the high momentum $\pi$ is misidentified as a muon, is
also indistinguishable from the signal decay and has a similar \nnout
shape. This contribution is fixed in the fit and the signal yield
difference, with and without the $B^-\to K_L^0 \pi^-$ component, of 5.5\%
is taken as a systematic uncertainty since {\tt GEANT3} poorly models
$K_L^0$ interactions with materials.

The not-yet-discovered process $B^{-}\to\mu^{-}\bar\nu_\mu \gamma$
with a soft photon can mimic the signal decay. To estimate the
uncertainty from this hypothetical background, we perform the fit with
this contribution fixed to half of the best upper limit
$\Br(B^{-}\to\mu^{-}\bar\nu_\mu \gamma)<3.4\times 10^{-6}$ at
90\%~C.L. by Belle~\cite{lnug} and take the difference of 6\% as the
systematic uncertainty.

Previous studies~\cite{Satoyama:2006xn,Aubert:2009ar} did not characterize these backgrounds in a
detailed manner, which could have led to a
substantial underestimation of the systematic uncertainties.

In the region $p_\mu^*>2.85$~\GeVc, where only
continuum events are present, we observe an almost linearly growing
data/fit difference with maximum deviation $\sim 20$\% at
$\nnout\sim1$. To estimate the uncertainty due to the
level of data/MC agreement in the \nnout variable, we 
rescale linearly with \nnout the continuum histograms used in the fit and refit,
obtaining a 15\% lower value of $R$.
For peaking components such as the signal \bmmumnub and the normalization
decay \bbpilmnub, we use the fit/data ratio in the region
$p_\mu^*<2.5$~\GeVc and apply it to the peaking components in the
signal-region histograms (\bmmumnub, \bbpilmnub and \brholnu).
Refitting produces an 11\% higher value of $R$. Simultaneously applying
both effects leads to only a 2\% shift in the refitted central value; thus, we
include the individual deviations as systematic uncertainties in the
continuum and signal peak descriptions.

In some cases, the signal muon and detected fraction of the particles from
the companion $B$-meson decay do not provide enough particles for an event 
to be identified as a $B$-meson decay and hence to be recorded. The efficiency for
recording these events is 84\% as calculated using MC, and we take 
the event-recording uncertainty to be half of the inefficiency (8\%) 
since it will be partially cancelled by taking the ratio with 
the normalization process \bbpilmnub.

The branching fraction of the normalization process \bbpilmnub is known 
with 3.4\% precision~\cite{PDG} and this is included as a systematic 
uncertainty.

The summary of the systematic uncertainties is shown in Table~\ref{table:syst}.
The total systematic uncertainty of 25\% is obtained by summing the
individual contributions discussed above in quadrature. 

\begin{table}[tbh]
  \caption{\label{table:syst} The summary of the systematic
    uncertainties for the branching fraction result.}
  \centering
  \begin{tabular}{cc}
    \hline
    Source & Uncertainty (\%)\\
    \hline
    \bbpilmnub form-factor & 0.9 \\
    \brholnu form-factor &  12 \\
    $B^-\to K_L^0 \pi^-$ & 5.5\\
    $B^{-}\to\mu^{-}\bar\nu_\mu \gamma$ & 6 \\
    Continuum shape & 15 \\
    Signal peak shape & 11 \\
    Trigger & 8 \\
    $\Br(\bbpilmnub)$ & 3.4\\
    \hline
    Total & 24.6 \\
    \hline
  \end{tabular}
\end{table}

Incorporating systematic uncertainties, the final branching
fraction for the signal decay is $\Br(\bmmumnub) = (6.46 \pm
2.22_\text{stat} \pm 1.60_\text{syst})\times10^{-7} = (6.46 \pm
2.74_\text{tot})\times10^{-7}.$ The accounted systematic uncertainties
reduce the fit statistical signal significance from 3.4 to 2.4
standard deviations. A confidence interval using a frequentist approach
based on Ref.~\cite{fc} is evaluated with systematic
uncertainties included and found to be $\Br(\bmmumnub) \in [2.9,
  10.7]\times 10^{-7}$ at the 90\% C.L., in agreement with the SM
prediction $\Br_\text{SM}(\bmmumnub) =( 3.80 \pm 0.31)\times 10^{-7}$.
  
In conclusion, as a result of an untagged search with the full Belle $\Upsilon(4S)$ data set,
we find a
2.4 standard deviation excess above background for the decay \bmmumnub, with a measured branching fraction
of $\Br(\bmmumnub) = (6.46 \pm 2.22_\text{stat} \pm
1.60_\text{syst})\times10^{-7}$ and a ratio of $\Br(\bmmumnub)/\Br(\bbpilmnub) =
(4.45 \pm 1.53_\text{stat} \pm 1.09_\text{syst}) \times 10^{-3}.$ The 90\% confidence interval for the
obtained branching fraction in the frequentist approach is
$\Br(\bmmumnub) \in [2.9, 10.7]\times 10^{-7}$. The forthcoming data
from the Belle~II experiment~\cite{Belle2} should further improve the
measurement.

\begin{acknowledgments}
We thank the KEKB group for the excellent operation of the
accelerator; the KEK cryogenics group for the efficient
operation of the solenoid; and the KEK computer group,
the National Institute of Informatics, and the 
PNNL/EMSL computing group for valuable computing
and SINET5 network support.  We acknowledge support from
the Ministry of Education, Culture, Sports, Science, and
Technology (MEXT) of Japan, the Japan Society for the 
Promotion of Science (JSPS), and the Tau-Lepton Physics 
Research Center of Nagoya University; 
the Australian Research Council;
Austrian Science Fund under Grant No.~P 26794-N20;
the National Natural Science Foundation of China under Contracts 
No.~10575109, No.~10775142, No.~10875115, No.~11175187, No.~11475187, 
No.~11521505 and No.~11575017;
the Chinese Academy of Science Center for Excellence in Particle Physics; 
the Ministry of Education, Youth and Sports of the Czech
Republic under Contract No.~LTT17020;
the Carl Zeiss Foundation, the Deutsche Forschungsgemeinschaft, the
Excellence Cluster Universe, and the VolkswagenStiftung;
the Department of Science and Technology of India; 
the Istituto Nazionale di Fisica Nucleare of Italy; 
National Research Foundation (NRF) of Korea Grants No.~2014R1A2A2A01005286, No.~2015R1A2A2A01003280,
No.~2015H1A2A1033649, No.~2016R1D1A1B01010135, No.~2016K1A3A7A09005603, No.~2016R1D1A1B02012900; Radiation Science Research Institute, Foreign Large-size Research Facility Application Supporting project and the Global Science Experimental Data Hub Center of the Korea Institute of Science and Technology Information;
the Polish Ministry of Science and Higher Education and 
the National Science Center;
the Ministry of Education and Science of the Russian Federation and
the Russian Foundation for Basic Research;
the Slovenian Research Agency;
Ikerbasque, Basque Foundation for Science and
MINECO (Juan de la Cierva), Spain;
the Swiss National Science Foundation; 
the Ministry of Education and the Ministry of Science and Technology of Taiwan;
and the U.S.\ Department of Energy and the National Science Foundation.
\end{acknowledgments}


\begin{thebibliography}{99}
\bibitem{conj}
    Charge-conjugate decays are implied throughout this paper.
\end{thebibliography}
\end{document}